# Novel Common Vehicle Information Model (CVIM) for Future Automotive Vehicle Big Data Marketplaces

Johannes Pillmann and
Christian Wietfeld
TU Dortmund University
Communication Networks Institute
{johannes.pillmann, christian.wietfeld}
@tu-dortmund.de

Adrian Zarcula and
Thomas Raugust
Volkswagen AG
VST/1 After Sales Technik
car2enterprise & Fahrzeug-EKG
{adrian.zarcula, thomas.raugust}
@volkswagen.de

Daniel Calvo Alonso
Atos Research & Innovation
Energy & Transport Market
daniel.calvo@atos.net

*Abstract*—Even though connectivity services have been introduced in many of the most recent car models, access to vehicle data is currently limited due to its proprietary nature. The European project *AutoMat* has therefore developed an open Marketplace providing a single point of access for brand-independent vehicle data. Thereby, vehicle sensor data can be leveraged for the design and implementation of entirely new services even beyond traffic-related applications (such as hyper-local traffic forecasts). This paper presents the architecture for a Vehicle Big Data Marketplace as enabler of cross-sectorial and innovative vehicle data services. Therefore, the novel Common Vehicle Information Model (CVIM) is defined as an open and harmonized data model, allowing the aggregation of brand-independent and generic data sets. Within this work the realization of a prototype CVIM and Marketplace implementation is presented. The two use-cases of local weather prediction and road quality measurements are introduced to show the applicability of the AutoMat concept and prototype to non-automotive applications.

## I. INTRODUCTION

Inside modern vehicles thousands of signals are processed. It is long ago that in-vehicle sensor signal's purpose was intended only to steering and safety related driving tasks. *Advanced Driver Assistance Systems (ADAS)* are pushing forward in new dimensions of passenger safety and comfort functionality as well as autonomous driving. Targeting the technologically high challenging autonomous driving, the number of installed sensors and complex signal processing units within vehicles is continuously growing. Millions of connected cars on Europe's streets create a huge potential for applications and services enabled through in-vehicle aggregated data. Today, this major potential is still locked since the automotive industry was not able to establish an open vehicle data based ecosystem. Inspired by the smartphone and mobile Internet industry this paper proposes an exchange platform for vehicle sensor data in form of a Vehicle Big Data Marketplace as enabler for cross-sectorial services.

Fig. 1 shows the processing chain for Vehicle Big Data aggregation. Vehicles are used as data miners to collect information about themselves and their environment using their built-in sensors. Gathered data is transfered into the cloud leveraging vehicle-to-cloud (V2C) communication systems. Due to the great heterogeneity of vehicle sensors the next and one of the most important steps in the processing chain is the harmonization of data. Brand-dependent sensors and different sample aggregation methods need to be lumped and brought to a common format. Communication from the vehicle into the cloud is often compressed and thinned out to save bandwidth and therefore needs to be enriched. This process ensures equality of measurements even from different sensors. Subsequently in the processing stage follows the aggregation of data over a whole fleet of vehicles. As one individual vehicle may not observe certain phenomenons, a whole fleet will as it increases the number of observed samples. The last step is the analysis of the aggregated data using Big Data techniques and methods, that extract information and knowledge to serve as basis for existing and novel cross-sectorial services.

This paper proposes a holistic marketplace architecture for Vehicle Big Data aggregation and exchange enabling service providers to ramp up large scale applications. Hereby, the CVIM data model, that harmonizes and generates brand-independent datasets, is defined.

## II. RELATED WORK

Sharing sensor information has been a topic of research in several fields of applications. In Wireless Sensor Networks (WSN), sensors exchange information in order to observe and survey large areas of interest (e.g. using temperature and humidity sensors for weather observation) [1]. Typical sensor networks are designed to be energy efficient and use-case specific. In contrast, vehicles are equipped with a variety of equipment and can be therefore called sensors on wheels. In the study [2], the collective perception potential of connected vehicles is evaluated. Enhanced and Extended Floating Car

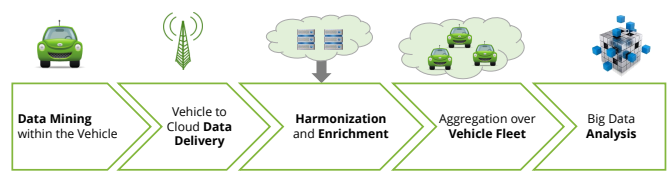

Fig. 1. Data Processing Chain for Vehicle Big Data Aggregation

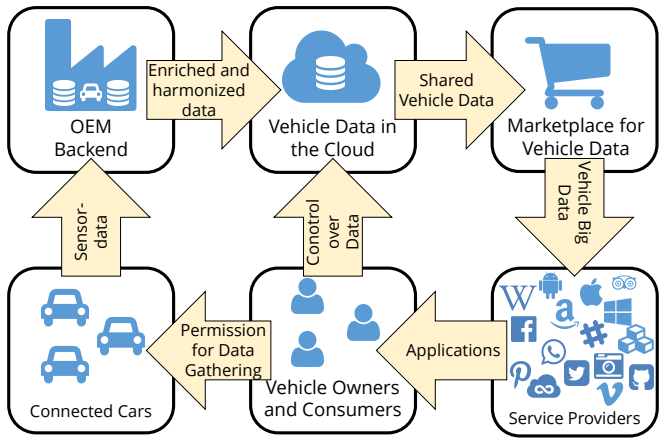

Fig. 2. Proposed Vehicle Big Data Marketplace Architecture and Data Flow

Data (xFCD), where vehicle sensor probes are sent from the vehicle to a central server, have also been introduced and analyzed in a variety of studies [3], [4]. In [5], the network capacity for car-to-cloud communication is analyzed.

In [6] and [7], challenges and solution approaches for the development and integration of data cloud services in the Internet of Things (IoT) environment were discussed and evaluated. Current car-to-cloud systems, like [8], are proprietary and do not allow unrestricted access to vehicle data. One first approach of a standardized vehicle sensor data access is provided by World Wide Web Consortium's (W3C) Vehicle Data standard [9]. The standard, which is currently still a working draft, focuses mainly on the in-vehicle interfaces and an evaluation of the car-to-cloud applicability must be yielded. In addition, the International Standardization Organization (ISO) enforces standardized vehicle data access interfaces using a neutral server instance with the focus on vehicle diagnosis services [10]. Under the coordination of ERTICO, the map making company *Here* is pushing forward its car-to-cloud standard *SensorIS* [11]. Also being under development at the moment of writing, the standard follows the concept of xFCD and focuses on time-series sensor measurements.

## III. MARKETPLACE ECOSYSTEM AND ARCHITECTURE

The architecture of the proposed Vehicle Big Data Marketplace is shown in Figure 2. It can be split into different components ordered around the central marketplace. Data gathering within vehicles is enabled by vehicle owners, drivers or other passengers. As the car is being driven, data is recorded inside the vehicle from 50 to 100 sensors, which are installed in modern vehicles. Sensors are the perception organs of vehicles and traditionally their detected information is used for ADAS, engine control and comfort functionality. With the advent of communication technology in cars, vehicles can be leveraged as rolling sensors to create major Big Data potentials.

After data has been mined inside the vehicle, it is transferred into the OEM backend. There, the raw data is enriched using OEM internal knowledge to transform brand-dependent into generic datasets. The OEM harmonizes proprietary information and brings them into the standardized CVIM format. This step is one of the most important to provide a mutual understanding of the data for all entities in the data processing chain. Due to the great variety of sensors no unified, open and non-proprietary vehicle data model exists. Therefore the CVIM as novel data model is introduced in the later sections.

Harmonized and enriched vehicle data is stored encrypted inside the vehicle user's personal cloud storage vault, depicted as *Vehicle Data in the Cloud* in Fig. 2. The cloud storage keeps data in separate and individual vaults for each vehicle user to enforce maximum possible privacy protection by design. Thereby, the *Vehicle Data in the Cloud* module follows the concept of citizen empowerment: the owner and creator of data stays in full control. Vehicle Data is only offered and shared on the Vehicle Big Data Marketplace, when the permission is actively given.

The marketplace plays a central role in the presented architecture. It serves as an exchange platform for vehicle data and fulfills the tasks of a data accumulator and intermediary between vehicle owners and service providers. Here, vehicle data from different sources merges together. On the supply side, vehicle owners offer data. Incoming data needs to be preprocessed and indexed to allow searching and an availability analysis. On the demand-side, service providers request data. They are able to search for available data and filter by region, age or quality. The Marketplace manages all offers and requests, collects the data from the individual vehicle user's cloud storage and combines it to Vehicle Big Data from a whole fleet.

In the last step of the processing chain, service providers build innovative applications based on Vehicle Big Data. These applications and services are offered in return to the vehicle users and other consumers.

## IV. COMMON VEHICLE INFORMATION MODEL (CVIM)

One key success factor of the Vehicle Big Data Marketplace is a unified and efficient data representation, harmonizing proprietary data as well as removing brand-specific information. As there are no standardized, non-proprietary data models, the novel CVIM has been developed within the AutoMat project.

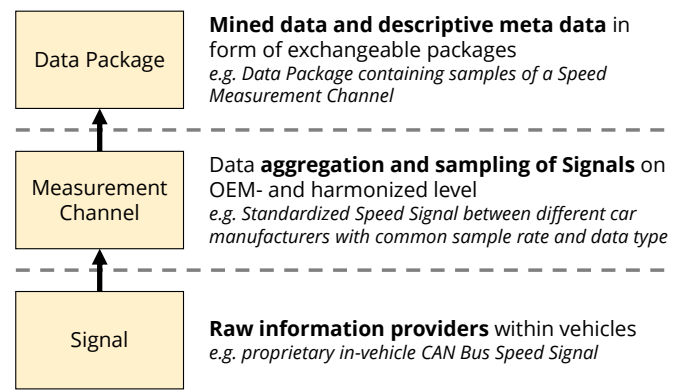

Fig. 3. Layered CVIM Data Model Architecture

The CVIM works on three separate layers (cf. Fig. 3). On the lowest layer, signals from inside vehicles are represented. These CVIM Signals are car manufacturer- and vehicle-specific and thereby proprietary. Their origin may be any source within the vehicle, e.g. a vehicle speed signal captured from the Controller Area Network (CAN) or On-Board Diagnostics (OBD) bus. The inner, middle layer consists of Measurement Channels. Measurement Channels define the common ground between signals and information originating from different vehicles or brands. They harmonize proprietary information into a standardized format by removing brand dependencies. On the top level, aggregated data is stored inside CVIM Data Packages. These packages are exchangeable messages, that can be transferred from vehicles via the respective OEM backend into the cloud storage.

### A. Signals

Signals are the fundamental information providers within the Vehicle Big Data processing chain. Their content is gathered from sensors, which are the perception organs of vehicles. Sensors observe the vehicle's environment by measuring physical phenomenons and chemical quantities and transferring them into sampled signals. Thereby, sensors form the basis of Connected Cars and vehicle data aggregation.

One major problem, which occurs at this stage and which the data model needs to take care of, is the great heterogeneous sensor landscape. The variety comes from the large number of different OEMs and their suppliers. Different companies use different sensors for the same observations. Within the same car manufacturer, sensors vary between models or even between cars of the same model due to changes of supplier. This implies differences in sample rate, resolution and biases. Additionally, sensor configurations vary with the equipped supplementary ADAS (e.g. parking assistant or lane departure warning). Therefore, two cars won't provide the same signals.

CVIM Signal layer solves this problem by harmonizing all possible types of signals. The description of basic observations and car status information are aligned and described in a brand-independent way. Within the AutoMat project 248 signals and information types were identified and defined. They form the basis for signal measurements on the next layer. Nevertheless, this CVIM Signal definition does not include the different sample rates and value ranges, which are still proprietary at this data model layer.

The data model concept follows the principle *"Every signal is optional, nothing is a must."* to address different vehicle sensor configurations. Hereby, CVIM allows vehicles with a minimum of sensors as well as fully-equipped premium segment cars to participate on the AutoMat Marketplace.

### B. Measurement Channels

Measurement Channels describe how data is aggregated and sampled from signals. Sensors are polled and evaluated inside vehicles several hundred to thousand times a second. With at least $50$ to $100$ different sensors installed, this produces an enormous amount of data. Due to limited storage and data transfer capacity, preprocessing of data samples is necessary. The following compressed data representations are supported by CVIM Measurement Channels:

**Time Series Measurement Channel (TSMC)**: TSMCs aggregate signal values over time. Hereby, the underlying signal's sample rate can be reduced by either keeping only one sample per interval (downsampling) or averaging over several samples. Fig. 4 shows an excerpt from a typically engine speed signal (sample rate $100\ Hz$) that is downsampled to meet the $1\ Hz$ sample rate of the engine speed TSMC.

This process aligns and harmonizes signals from different OEMs and is therefore strongly necessary. The CVIM concept allows multiple Measurement Channels originating from one signal to enable higher-quality data channels, e.g. while most of the vehicles deliver the engine speed only with $1\ Hz$ sample rate, there may be a few vehicles delivering the same signal with $100\ Hz$. In this case, cars deliver data ino the *high quality* TSMC and in addition to the *standard quality* TSMC. An analysis of the error of data reduction is given in [12], which may guide in choosing the correct sample rate for signals.

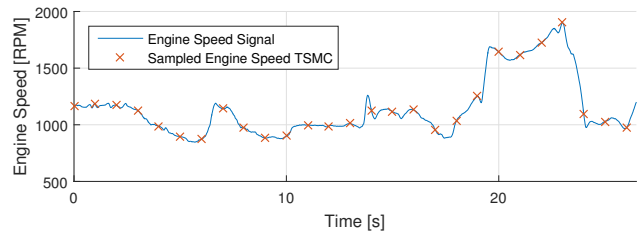

Fig. 4. Excerpt from a typical engine speed signal with downsampling for the engine speed Time-Series Measurement Channel

**Histogram Measurement Channel (HMC):** Contrary to the TSMC in HMC no samples are dropped due to down-sampling. Instead, the time-domain of signals is removed by only keeping the value distribution in form of histograms. Histograms do not require neither high implementation nor complex preprocessing within the car but only a fraction of the network capacity compared against TSMC. CVIM supports multidimensional histograms, where each dimension represents one signal. Fig. 5 shows an exemplarily two-dimensional HMC of vehicle speed and engine speed signals. A typical use case for histogram analysis are wear predictions of vehicle components.

**Geo-based Histogram Measurement Channel (GHMC):** GHMCs are similar to HMC as GHMCs also store signal values distributions in form of histograms. In addition, a geographic resolution is added, describing the size of area in which the histogram was recorded. Hereby, GHMCs allow spatial correlation of histograms. Fig. 6

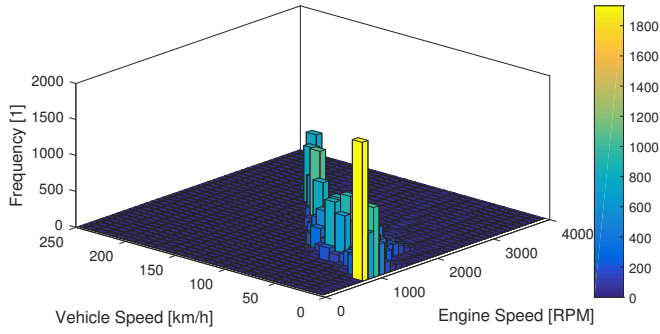

Fig. 5. Two-dimensional Histogram Measurement Channel of Vehicle Speed and Engine Speed Signals

shows an GHMC example of a whiper signal analyzed along the vehicle's route.

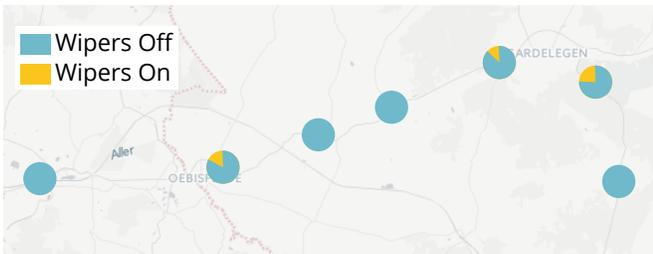

Fig. 6. Geo-based Histogram Measurement Channel of a Wiper-Signal

*C. Data Packages*

CVIM Data Packages contain the actual data of measurement channels. They are the exchangeable data structure used to transfer data between all entities in the AutoMat concept. Data Packages are split into descriptive meta-data and the aggregated data (cf. Fig. 7). Aggregated data contains samples from measurement channels as introduced in the previous sections, whereas meta-data describes those data samples in order to help the marketplace identify, search and classify relevant data for service providers without analyzing whole datasets. Next to beginning and ending timestamps, the vehicle's mileage and geographic boundaries are included. Statistic properties as minimum, maximum and average give a short preview of the data. In addition, OEMs ensure quality, completeness and correctness of data by providing a signature, checksums and sequence numbers. Ownership attributes contain information about the data's copyright stakeholders as well as the privacy level of the data. Fig. 8 depicts the prototype of the cloud data storage, showing the meta-data of a speed data package of one test vehicle.

V. THE VEHICLE BIG DATA MARKETPLACE

The Marketplace is a mediator platform that offers access to brand independent vehicle information, which is represented using the CVIM. It ensures the flow of data from the vehicle owners' private cloud storage to the service providers' backends to enable the creation of new services and applications.

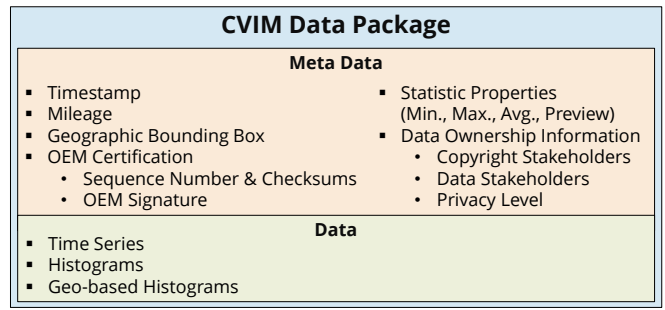

Fig. 7. Structure and Content of CVIM Data Packages

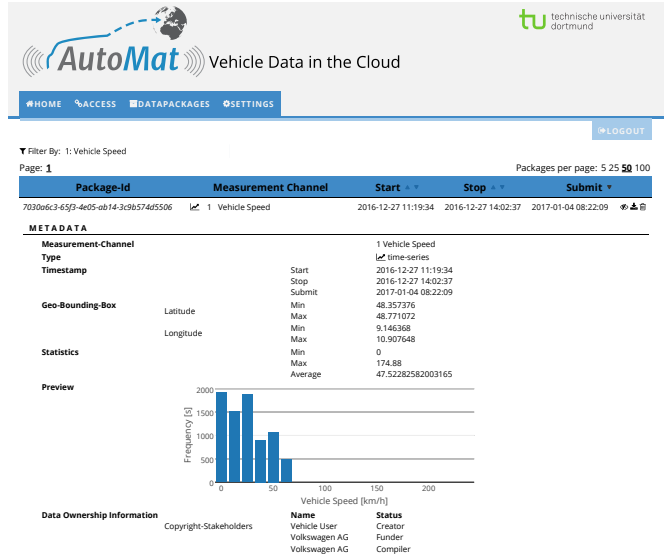

Fig. 8. Cloud Data Storage Prototype Showing one CVIM Data Package

The marketplace includes key features to cover the complete life-cycle of Vehicle Big Data services as shown in Fig. 9. All features are exposed via two different interfaces: one REST (REpresentational State Transfer) interface, specified in a machine-readable format for seamless integration with services providers' existing infrastructures, and one responsive cross-platform front-end for human-friendly interaction.

*A. Data Exploration and Discovery*

For exploration and discovery purposes service providers are able to browse through the *CVIM Catalog*, which contains available CVIM Measurement Channels and supports the identification of relevant information for new services. Indexing and analysis of the meta data of CVIM Data Packages provide statistics of the amount and quality of available datasets and sources, e.g. update frequency, ge-ographical and time distribution. Hereby, service provider identify appropriate data for their services.

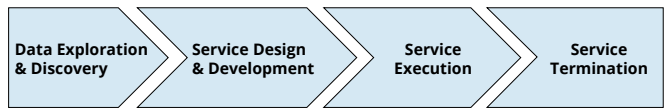

Fig. 9. Vehicle Big Data Service Life-cycle Phases

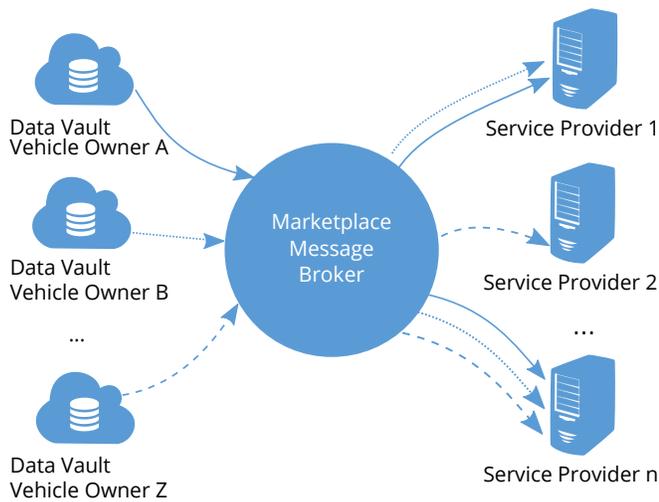

Fig. 10. Marketplace Message Broker That Manages Data Flows Between Private Data Storage Vaults and Service Providers

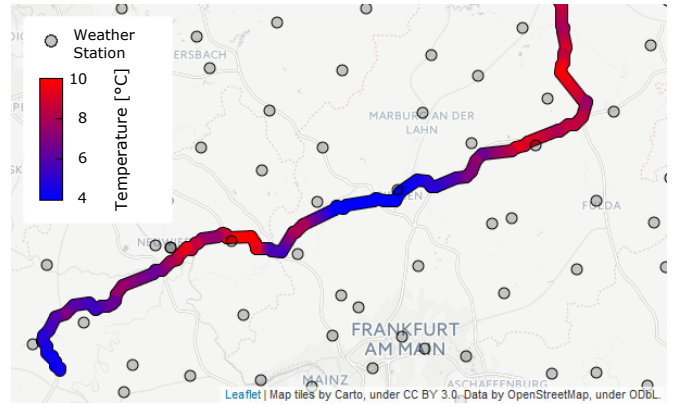

Fig. 11. Temperature Trajectory of a Test Vehicle With Surrounding Federal German Meteorological Observation Stations

### B. Service Design and Development

During the service design and development phase the AutoMat Marketplace supports service providers with vehicle test data and a Software Development Kit, which enables simple access to the marketplace functionalities and interfaces as well as rapid integration into the application backends.

For the initialization of data flows, access agreements between service providers and data owners are set up and managed by the marketplace. The marketplace integrates a message broker as part of its internal architecture as shown in Fig. 10. This message broker handles the routing of all data flows according to the agreements from the vehicle owners private data vaults to the service providers.

### C. Service Execution

During run-time of the service, the marketplace transfers data from the vehicle owners' personal cloud storage vaults to the service providers according to the arranged agreements. The following data retrieval approaches are supported:

**Pull Mode**: In pull mode, service providers perform queries to receive data from the marketplace. The Marketplace validates, whether new data is available, retrieves and accumulates data from all cloud storage vaults, which is delivered afterwards. This approach features the data retrieval of one whole vehicle-fleet in one request.

**Push Mode**: In the alternative push mode, the cloud storage forwards incoming CVIM Data Packages via the marketplace's message broker to the service providers according to access agreements. This approach minimizes the delay between observation and processing and enables highly fresh data; the accumulation for all individual vehicles to fleet data has to be performed by the service provider.

### D. Service Termination

At any time, vehicle owners may decide to stop sharing their data, canceling the corresponding agreements. The data flow from the corresponding storage vault is stopped. On the other side, service providers can as well end their data subscriptions, where data flow from all sources is terminated and unregistered from the message broker.

## VI. USE CASES

When it comes to gathering and evaluating vehicle data, the most prominent use case is the collection of current and the prediction of future traffic. Additionally, the marketplace explicitly addresses cross-industry applications. Within the scope of the AutoMat project the two example use cases of hyper local weather prediction and improved road quality measurement enabled by vehicle big data.

### A. Hyper-Local Weather Prediction

The provision of meteorological data from observation networks with a high spatial resolution remains a challenge in weather prediction. As a cost-benefit compromise, the inter-site distance between observation stations is usually multiple kilometers. Thus, regional weather phenomenons, e.g. a local hail storm, are often unnoticed by official meteorological observation stations. Fig. 11 shows a temperature trajectory of one test vehicle with the surrounding meteorological weather observation stations of the federal German weather service DWD[1]. The vehicle as a moving sensor closes the gaps between those stations. Next to temperature, vehicles also provide humidity, dew-point, fog and rain detection as well as several other information. Studies as [13] have evaluated the benefits of crowd-sourced temperature measurements for stationary *netatmo* consumer weather sites. Therefore, classical weather models enriched with Vehicle Big Data will lead to more accurate and more regional weather predictions. With millions of cars on Europe's streets precise spatial coverage also in currently sparsely observed areas is possible.

[1] https://www.dwd.de/DE/leistungen/klimadatendeutschland/statliste/statlex_html.html?view=nasPublication&nn=16102

## B. Road Quality

Monitoring the road quality is a crucial aspect of road maintenance. Damages and potholes affect the security of road traffic negatively. Furthermore, the driving comfort is reduced and the higher rolling resistance leads to higher emissions. Detection and remedy has therefore high priority. Many solutions have been proposed to check the road quality - most of them make use of expensive sensors or special equipped vehicles. Low cost collaborative solutions make use of accelerometers built into smartphones (e.g. [14]). While smartphones are widely available and easy to deploy, the accuracy and quality of road quality detection is still limited. With the availability of vehicle sensor data the accuracy can be increased significantly. In addition to acceleration, suspension height, running gear signals and further vehicle properties are available and can also be taken into deeper analysis. With the high number of vehicles the significance of measurements can be increased and damages or potholes can be detected more accurately.

## VII. CONCLUSION

In this paper a concept for an holistic architecture for vehicle big data aggregation is presented. The proposed system enables providers of services and applications to access vehicle data via a single point of access - the AutoMat marketplace. The marketplace handles vehicle fleet accumulation and contract handling with vehicle owners while preserving privacy rights of all data stakeholders. The proposed unified data format *CVIM* solves the tasks of harmonizing proprietary and brand-dependent sensor data and information. CVIM allows different levels of quality as well as different data aggregation types in form of time-series and (geo-based) histograms.

As shown at the examples of weather prediction and road quality monitoring, the proposed AutoMat platform enables exploitation of vehicle big data for innovative and cross-sectorial applications and services.

## ACKNOWLEDGMENT

This work has been conducted within the *AutoMat* (Automotive Big Data Marketplace for Innovative Cross-sectorial Vehicle Data Services) project, which received funding from the European Union's Horizon 2020 (H2020) research and innovation programme under the Grant Agreement no 644657, and has been supported by Deutsche Forschungsgemeinschaft (DFG) within the Collaborative Research Center SFB 876 Providing Information by Resource-Constrained Analysis, project B4.